\documentclass[twocolumn,prl,showpacs,amsfonts,amsmath,assymb,eufrak]{revtex4}

\usepackage{epsfig}
\def\eq#1\en{\begin{equation}#1\end{equation}}  
\def\eqa#1\ena{\begin{align}#1\end{align}}
\def\eqg#1\eng{\begin{gather}#1\end{gather}}
\newcommand{\lb}[1]{\label{e:#1}}
\newcommand{\rlb}[1]{\eqref{e:#1}} 
\newcommand{\nl}{\notag\\}

\newcommand{\bkt}[1]{\left\langle#1\right\rangle}
\newcommand{\sbkt}[1]{\langle#1\rangle}
\newcommand{\bbkt}[1]{\bigl\langle#1\bigr\rangle}

\newcommand{\sumtwo}[2]%
{\mathop{\sum_{#1}}_{#2}}
\newcommand{\sumthree}[3]%
{\mathop{\mathop{\sum_{#1}}_{#2}}_{#3}}
\newcommand{\sumfour}[4]%
{\mathop{\mathop{\mathop{\sum_{#1}}_{#2}}_{#3}}_{#4}} 
\newcommand{\prodtwo}[2]%
{\mathop{\prod_{#1}}_{#2}}
\newcommand{\mintwo}[2]%
{\mathop{\min_{#1}}_{#2}}
\newcommand{\maxtwo}[2]%
{\mathop{\max_{#1}}_{#2}}
\newcommand{\maxthree}[3]%
{\mathop{\mathop{\max_{#1}}_{#2}}_{#3}}
\newcommand{\limtwo}[2]%
{\mathop{\lim_{#1}}_{#2}}
\newcommand{\suptwo}[2]%
{\mathop{\sup_{#1}}_{#2}}
\newcommand{\supthree}[3]%
{\mathop{\mathop{\sup_{#1}}_{#2}}_{#3}}
\newcommand{\supfour}[4]%
{\mathop{\mathop{\mathop{\sup_{#1}}_{#2}}_{#3}}_{#4}} 
\newcommand{\inftwo}[2]%
{\mathop{\inf_{#1}}_{#2}}
\newcommand{\infthree}[3]%
{\mathop{\mathop{\inf_{#1}}_{#2}}_{#3}}
\newcommand{\inffour}[4]%
{\mathop{\mathop{\mathop{\inf_{#1}}_{#2}}_{#3}}_{#4}} 

\newcommand\calE{{\cal E}}

\newcommand\calM{{\cal M}}

\newcommand\calT{{\cal T}}

\newcommand{\bsp}{\boldsymbol{p}}

\newcommand{\bsr}{\boldsymbol{r}}

\newcommand{\HE}{H}
\newcommand{\HM}{\tilde{H}}
\newcommand{\WE}{W}
\newcommand{\WM}{\tilde{W}}
\newcommand{\ZE}{Z}
\newcommand{\ZM}{\tilde{Z}}
\newcommand{\Ga}{\Gamma}
\newcommand{\Up}{\Upsilon}
\newcommand{\TmE}{\calT^\mathrm{ms}}
\newcommand{\TmM}{\tilde{\calT}^{\mathrm{ms}}_{\Ga}}
\newcommand{\TfE}{\calT^{\mathrm{fb}}_{\Up'}}
\newcommand{\TfM}{\tilde{\calT}^\mathrm{fb}}
\newcommand{\TE}{\calT_{\Up'}}
\newcommand{\TM}{\tilde{\calT}_{\Ga}}

\newcommand{\TmMmu}{\tilde{\calT}^{\mathrm{ms}}_{\mu}}
\newcommand{\TfEmu}{\calT^{\mathrm{fb}}_{\mu}}
\newcommand{\TEmu}{\calT_{\mu}}
\newcommand{\TMmu}{\tilde{\calT}_{\mu}}
\newcommand{\rhoz}{\bar{\rho}_0}
\newcommand{\rhoM}{\tilde{\rho}_0}
\newcommand{\rhoE}{\rho_0}
\newcommand{\trho}{\tilde{\rho}}
\newcommand{\bS}{\bar{S}}
\newcommand{\tS}{\tilde{S}}
\newcommand{\brho}{\bar{\rho}}

\newcommand{\para}[1]{{\em #1}\/.---}
\newcommand{\midskip}{\vspace{3pt}}

\newcommand{\hal}[1]{\relax}

\begin{document}
\title{
Unified Jarzynski and Sagawa-Ueda relations for Maxwell's demon
}

\author{Hal Tasaki}
\affiliation{
Department of Physics, Gakushuin University, 
Mejiro, Toshima-ku, Tokyo 171-8588, Japan}

\date{\today}

\begin{abstract}
By using Newtonian mechanics, we construct a general model of Maxwell's demon, a system in which the engine and the memory interact only through the exchange of information.
We show that the Jarzynski relation and the two Sagawa-Ueda relations hold simultaneously, and argue that they are the unique triplet which has a natural decomposition property.
The uniqueness provides a strong support to the assertion that the mutual information is the key quantity.
\end{abstract}


\maketitle
Recently there has been a considerable renewed interest in the problem of Maxwell's demon \cite{Maxwell,DemonBook,MaruyamaNoriVedral}.
Based on progress in the twentieth century \cite{DemonBook,MaruyamaNoriVedral,Szilard,Brillouin,Landauer,Bennett} which revealed the essential role of information, and more recent progress in nonequilibrium physics \cite{Jarzynski97,Crooks,Seifert12} in particular the Jarzynski relation and similar results, mathematically refined theories related to demon have been developed \cite{TouchetteLloyd,Piechocinska,KawaiParrondoVandenBroeck,KimQian}.
In particular Sagawa and Ueda have derived a series of general and exact results \cite{%
SagawaUeda2008,SagawaUeda2009,SagawaUeda2010,SagawaUeda2012B,SagawaUeda2012v1,SagawaUeda2012,SagawaUeda2013%
} which shed light on the essence of Maxwell's demon (or, more generally, systems where measurement and feedback are essential) and  suggest a fundamental role played by mutual information.

Imagine a (probably small) thermodynamic system, such as the Szilard engine \cite{Szilard,DemonBook,MaruyamaNoriVedral}, which is subject to measurement and feedback.
It is well-known that such an ``engine'' may produce more work than that is allowed by the second law of thermodynamics.
Then the key question is how much extra work is needed to operate the device, which may be called a demon, that realizes the measurement/feedback.
It is believed that in principle such a device can be made as efficient as possible so that to waste less and less energy, except for a single component, the ``memory'', which stores the information about the engine \cite{DemonBook,MaruyamaNoriVedral,Bennett}.

This motivates us to study, in the present paper, a composite system of simultaneously evolving ``engine'' and ``memory'' \footnote{%
To our knowledge such a composite system was first studied by Sagawa and Ueda  in \cite{SagawaUeda2012v1}.
} that behaves (almost) as a normal physical system as a whole.
By constructing such a system within classical mechanics, we can analyze the flow of energy and entropy completely, and realize a situation in which the engine and the memory interact only thorough the exchange of information.
This construction provides a definite and most strict criterion of which system should be regarded as a Maxwell's demon, provided that we restrict ourselves to a classical system and allow an external agent who operates on the system.

We then prove the Jarzynski relation and the two Sagawa-Ueda relations which involve mutual information, recovering the known results in the unified setting.
These relations yield the standard and the extended second laws as usual.
More importantly we show that the above three relations are the unique triplet of integral fluctuation relations which satisfies a natural decomposition property.
This uniqueness provides a strong support to the assertion that mutual information plays a fundamental role in the problem of Maxwell's demon \cite{%
SagawaUeda2008,SagawaUeda2009,SagawaUeda2010,SagawaUeda2012B,SagawaUeda2012v1,SagawaUeda2012,SagawaUeda2013%
}.

We believe that our results do not only complete the project of Sagawa and Ueda (for a classical \footnote{%
See \cite{FunoWatanabeUeda} for a somewhat similar treatment of quantum systems.
} non-autonomous demon), but also can be a crucial guide in further studies of a variety of systems which share certain aspects of Maxwell's demon \cite{%
ToyabeSagawa,MandalJarzynski,HorowitzSagawaParrondo,StrasbergSchallerBrandesEsposito1,MandalQuanJarzynski,StrasbergSchallerBrandesEsposito%
}.

\midskip
\para{Setup and time-evolution}%
We consider a system of classical particles which consists of two subsystems, the {\em engine}\/ and the {\em memory}\/.
The state of the engine is collectively denoted as $\Ga=(\bsp_1,\ldots,\bsp_N,\bsr_1,\ldots,\bsr_N)\in\calE$, the state of the memory as $\Up=(\tilde\bsp_1,\ldots,\tilde\bsp_{\tilde{N}},\tilde\bsr_1,\ldots,\tilde\bsr_{\tilde{N}})\in\calM$, and the state of the whole system as $(\Ga,\Up)\in\calE\times\calM$.
We also write $d\Ga=\prod_{j=1}^Nd^3\bsp_jd^3\bsr_j$ and $d\Up=\prod_{j=1}^{\tilde{N}}d^3\tilde\bsp_jd^3\tilde\bsr_j$.

Physically speaking the ``engine'' consists of the main body of the engine and a heat bath associated with it, and the ``memory'' consists of the memory itself and another bath.
We have prepared separate heat baths so that to precisely trace the interaction between the engine and the memory.
In what follows we shall not explicitly mention about the baths, but we always understand that they are included in the engine or the memory.

\begin{figure}
\centerline{\epsfig{file=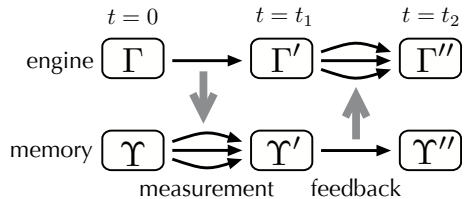,width=6cm}}
\vspace{-0.2cm}
\caption[dummy]{
Schematic picture of the time-evolution.
}
\vspace{-0.3cm}
\label{f:EM}
\end{figure}

Both the engine and the memory are isolated from the external world, and evolve according to the Newtonian mechanics.
We assume however that the engine and the memory are operated by an outside agent, and their Hamiltonians are varied in time according to protocols which are fixed in advance.
The protocols are designed so that to realize {\em measurement}\/ in the first period with $t\in[0,t_1]$, and {\em feedback}\/  (and {\em memory erasure}\/) in the second period with $t\in[t_1,t_2]$.
See Fig.~\ref{f:EM}.
We denote by $\HE$ and $\HM$ the Hamiltonians of the engine and the memory, respectively, at the initial time $t=0$.

In the period $[0,t_1]$ of measurement, the engine evolves according to a fixed protocol.
We denote by $\TmE:\calE\to\calE$ the corresponding time-evolution map (which brings the state at $t=0$ to that of $t=t_1$).
The memory also evolves according to a protocol, but the choice of the protocol is affected by the state of (the ``main body'' of) the engine in $[0,t_1]$.
Mathematically we can assume that the protocol is specified by the state of the engine at $t=0$, which we write $\Ga$.
The corresponding time-evolution map is $\TmM:\calM\to\calM$.
We assume that the Hamiltonian $\HM'$ of the memory at $t=t_1$ is independent of $\Ga$.
The idea is that the state, not the Hamiltonian, of the memory at $t=t_1$ records information about $\Ga$.

In the period  $[t_1,t_2]$ of feedback (and erasure), the engine and the memory switch their roles.
The engine now evolves according to a protocol which depends on the state of the memory at $t=t_1$, which we write $\Up'$.
This dependence represents the feedback \footnote{%
We are, in a sense, assuming that the agent itself has no memory.
In the period of feedback, it does not remember which protocol was chosen in the period of measurement.
}.
The time-evolution map is denoted as $\TfE:\calE\to\calE$.
The memory evolves according to a fixed protocol.
We suppose that the time-evolution $\TfM:\calM\to\calM$ finally erases the information stored in the memory \footnote{%
By only using the deterministic Newtonian mechanics it is impossible to completely delete the information about the state at $t=t_1$.
We expect however that the recovery can be made practically impossible by designing a proper dynamics since the heat bath contains many particles whose motion can be complicated.
Anyway we shall not make use of any assumptions about memory erasure in the derivation of the results.
}.
We assume that the whole process is cyclic in the sense that the Hamiltonians of the system and the memory at $t=t_2$ return to $\HE$ and $\HM$, respectively \footnote{%
The assumption of cyclicity is not at all essential.
If the process is not cyclic, one should include the differences of the initial and the final free energies in the main equalities  \rlb{J}, \rlb{SUE}, and \rlb{SUM}.
}.

Finally we denote by $\TE=\TfE\circ\TmE$ and $\TM=\TfM\circ\TmM$ the time-evolution maps of the engine and the memory, respectively, for the whole time interval.

\midskip
\para{Basic properties of the system}%
Recall that the Liouville theorem is valid when the Hamiltonian changes according to a fixed protocol.
Thus each of the maps $\TmE$, $\TmM$ (with any fixed $\Ga$), $\TfE$ (with any fixed $\Up'$), and $\TfM$ preserves the phase space volume.
We further assume that each of them is a one-to-one map \footnote{
The one-to-one property does not follow automatically and should be assumed.
In the present context, it means that measurement and feedback are associated with some errors.
The similar problem with no errors (in which the one-to-one property no longer holds) can also be treated, both in classical and quantum settings.
}.

Let $(\Ga,\Up)$ be the state at $t=0$, and denote the corresponding states at $t=t_1$ as $(\Ga',\Up')$, and at $t=t_2$ as $(\Ga'',\Up'')$, i.e.,
\eqg
\Ga'=\TmE(\Ga),\quad\Up'=\TmM(\Up),
\\
\Ga''=\TfE(\Ga'),\quad\Up''=\TfM(\Up').
\eng
We remark that the map from $(\Ga,\Up)$ to $(\Ga',\Up')$ is one-to-one.
To see this, take an arbitrary $(\Ga',\Up')$, and note that $\Ga=(\TmE)^{-1}(\Ga')$ uniquely determines $\Ga$, and then $\Up=(\TmM)^{-1}(\Up')$ uniquely determines $\Up$.
The map from $(\Ga,\Up)$ to $(\Ga',\Up')$ also preserves the phase space volume since both $\TmE$ and $\TmM$ (for a fixed $\Ga$) do.
Since the same observation is valid for the map from $(\Ga',\Up')$ to  $(\Ga'',\Up'')$, we find that the map from the initial state $(\Ga,\Up)$ to the final state $(\Ga'',\Up'')$ is also one-to-one and preserves the phase space volume.

We believe that we have defined an ideal class of mechanical systems which captures the essence of Maxwell's demon (or, more precisely, Szilard's interpretation of the demon) in the following two senses.

First the engine and the memory are carefully designed so that to interact with each other only through the ``exchange of information''.
Since the engine and the memory evolve separately as isolated systems, they exchange energy only with the external agent, and not with each other.
Moreover the fact that the time-evolutions of the engine and the memory separately preserve their phase space volumes implies that there are no mechanical exchange of entropy between them.
The only interaction between the engine and the memory arises from the choice of the protocol by the external agent.

Secondly the time-evolution of the whole system (but not that of the engine or the memory) is one-to-one and preserves the phase space volume.
This means that our system, as a whole, behaves (almost) as a normal Newtonian mechanical system.

\midskip
\para{Main results}%
We assume that at $t=0$ the state $(\Ga,\Up)$ is drawn from the probability distribution $\rhoz(\Ga,\Up)=\rhoE(\Ga)\,\rhoM(\Up)$, where
\eq
\rhoE(\Ga):=\frac{e^{-\beta\HE(\Ga)}}{\ZE},\quad
\rhoM(\Up):=\frac{e^{-\beta\HM(\Up)}}{\ZM}
\lb{can}
\en
are the canonical distributions.

For any function $F(\Ga,\Up)$ of the initial state $(\Ga,\Up)$, we define its average as
\eq
\sbkt{F(\Ga,\Up)}:=\int d\Ga d\Up\,F(\Ga,\Up)\,\rhoz(\Ga,\Up).
\en

Let us define (with $\Up'$ being a free variable)
\eq
\trho(\Up'|\Ga):=\int d\Up\,\delta\bigl[\Up'-\TmM(\Up)\bigr]\,\rhoM(\Up),
\lb{Up'cond}
\en
which is the probability density to get $\Up'$ in the memory at $t=t_1$ given the condition that the engine was in $\Ga$ at $t=0$.
We also write the unconditioned probability density as
\eq
\trho(\Up'):=\int d\Ga\,\trho(\Up'|\Ga)\,\rhoM(\Ga).
\lb{Up'prob}
\en
We then define the mutual information function as
\eq
I(\Ga,\Up'):=\log\frac{\trho(\Up'|\Ga)}{\trho(\Up')},
\en
whose average
\eqa
\bar{I}&:=\bbkt{I(\Ga,\TmM(\Up))}
\nl&
=\int d\Ga d\Up'\,\trho(\Up'|\Ga)\,\rhoM(\Ga)
\log\frac{\trho(\Up'|\Ga)}{\trho(\Up')}\ge0
\lb{Ibar}
\ena
is the mutual information between the state of the engine at $t=0$ and that of the memory at $t=t_1$ \footnote{
The final expression in \rlb{Ibar} can be easily derived by proceeding as in \rlb{eng1}.
}.

We also define
\eqg
\WE(\Ga,\Up):=\HE(\Ga)-\HE(\calT_{\TmM(\Up)}(\Ga)),
\lb{WE}
\\
\WM(\Ga,\Up):=\HM(\Up)-\HM(\TM(\Up)),
\lb{WM}
\eng
which are the works done by the engine and by the memory, respectively, to the agent during the whole process.

Our main results are the three equalities
\eqg
\bbkt{e^{\beta\{\WE(\Ga,\Up)+\WM(\Ga,\Up)\}}}=1,
\lb{J}
\\
\bbkt{e^{\beta\WE(\Ga,\Up)-I(\Ga,\Up')}}=1,
\lb{SUE}
\\
\bbkt{e^{\beta\WM(\Ga,\Up)+I(\Ga,\Up')}}=1,
\lb{SUM}
\eng
where $\Up'$ in the expectations should be replaced by $\TmM(\Up)$.

Eq. \rlb{J} is nothing but the original Jarzynski relation \cite{Jarzynski97} applied to the whole system.
The relations \rlb{SUE} and $\rlb{SUM}$ are the Sagawa-Ueda relations for feedback \cite{SagawaUeda2010} and for measurement \cite{SagawaUeda2012}, respectively.
See also \cite{ItoSagawa,FunoWatanabeUeda}.

We recall that, combined with the Jensen inequality $e^{\sbkt{F}}\le\sbkt{e^F}$, the relations \rlb{J}, \rlb{SUE}, and \rlb{SUM} lead to the standard second law for the whole system
\eq
\bbkt{\WE(\Ga,\Up)+\WM(\Ga,\Up)}\le0,
\lb{2nd}
\en
the generalized second law for the engine \cite{SagawaUeda2008}
\eq
\bbkt{\WE(\Ga,\Up)}\le\bar{I}/\beta,
\lb{2ndE}
\en
and that for the memory \cite{SagawaUeda2009}
\eq
\bbkt{\WM(\Ga,\Up)}\le-\bar{I}/\beta,
\lb{2ndM}
\en
respectively.
As is well understood by now, the engine may operate beyond the limit of the standard second law as in \rlb{2ndE}, but one must instead supply extra work to the memory as in \rlb{2ndM}.
Note that the inequalities \rlb{2nd}, \rlb{2ndE}, and \rlb{2ndM} are simultaneously saturated in a system of the Szilard engine and the standard (theoretical) memory consisting of a single gas molecule \footnote{%
To be rigorous we have to add small errors to the system so as to make it satisfy the conditions of the present work.
}.
See \cite{HorowitzParrondo} for the condition of saturation for the engine.

Note that 
the decomposition of the total work
\eq
\beta(\WE+\WM)=\{\beta\WE-I\}+\{\beta\WM+I\}
\lb{dec}
\en
has a remarkable property that the quantity in the left-hand side and the two quantities in the right-hand side simultaneously satisfy integral fluctuation relations (i.e., $\sbkt{e^F}=1$).
We call such a decomposition a Sagawa-Ueda decomposition \footnote{%
More abstractly, a Sagawa-Ueda decomposition is a special case of a decomposition $A=B+C$ with the property that $\sbkt{e^A}=\sbkt{e^B}=\sbkt{e^C}=1$.
Although we still do not know what this exactly implies, we remark that it is a highly nontrivial property which can hardly be realized accidentally.
The same type of decomposition is found in a driven nonequilibrium system where one decomposes the total entropy production into the sum of the ``house-keeping'' part and the remainder.
See \cite{HatanoSasa,SpeckSeifert} and also \cite{EspositoVandenBroeck}
} since, to our knowledge, the similar notion first appeared in \cite{SagawaUeda2012v1}.
See also \cite{ItoSagawa,SagawaUeda2013,FunoWatanabeUeda}.

More importantly, we will show that \rlb{dec} is the unique Sagawa-Ueda decomposition of the total work in the following sense.
As we shall see in the derivation, we have
\eq
\bbkt{e^{\beta\WE(\Ga,\Up)-X(\Ga,\Up)}}=1,\quad
\bbkt{e^{\beta\WM(\Ga,\Up)+Y(\Ga,\Up)}}=1.
\lb{XY}
\en
for several different $X$ or $Y$ including $Y=0$.
But if we further demand that $X=Y$ so that \rlb{XY} corresponds to a decomposition of the total work, our choice is essentially unique (in a certain weak sense to be read off from the derivation) and we have $X=Y=I(\Ga,\Up')$.

This uniqueness is a strong support for the assertion by Sagawa and Ueda that the mutual information is the key to understand Maxwell's demon and other problems where measurement and feedback are essential \cite{%
SagawaUeda2008,SagawaUeda2009,SagawaUeda2010,SagawaUeda2012B,SagawaUeda2012v1,SagawaUeda2012,SagawaUeda2013%
}.

\begin{figure}
\centerline{\epsfig{file=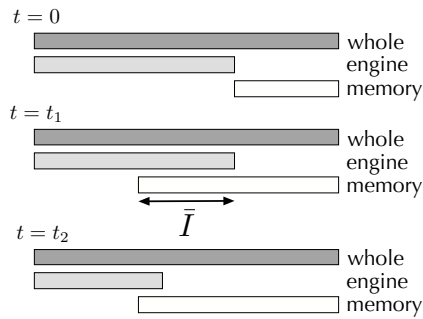,width=5.5cm}}
\vspace{-0.2cm}
\caption[dummy]{
The entropies in the initial state ($t=0$), after measurement ($t=t_1$), and after feedback ($t=t_1$).
}
\vspace{-0.3cm}
\label{f:SI}
\end{figure}

\midskip
\para{Entropies and mutual information}%
It is illuminating to consider how the entropies behave in the processes of measurement and feedback.
See Fig.~\ref{f:SI}.
Let $\brho_t(\Ga,\Up)$ be the probability distribution of the state of the whole system at time $t$.  
(Note that $(\Ga,\Up)$ is used as free variables, not as the initial state.)
The Shannon entropies \footnote{%
One should note that the entropies include those of the heat baths.
} at time $t$ of the whole system, the engine, and the memory are
$
\bS(t):=
-\int d\Ga d\Up\,\brho_t(\Ga,\Up)\,\log\brho_t(\Ga,\Up)
$,
$S(t):=-\int d\Ga\,\rho_t(\Ga)\,\log\rho_t(\Ga)
$,
and
$
\tS(t):=-\int d\Up\,\trho_t(\Up)\,\log\trho_t(\Up)
$,
respectively, with $\rho_t(\Ga):=\int d\Up\,\brho_t(\Ga,\Up)$ and $\trho_t(\Up):=\int d\Ga\,\brho_t(\Ga,\Up)$.

Note that $\bS(0)=S(0)+\tS(0)$ because the initial probability distribution splits.
Since the time-evolution of the whole system is always one-to-one and preserves the phase space volume, the entropy of the whole system is conserved, i.e., $\bS(t)=\bS(0)$ for any $t\in[0,t_2]$.

In the period $[0,t_1]$ of measurement, the entropy of the engine does not change since the time-evolution is simply that of an isolated system.
In particular we have $S(t_1)=S(0)$.
For each fixed $\Ga$, the time-evolution $\TmM$ of the memory also preserves the entropy.
Since the probability distribution $\trho_{t_1}(\Up)$ is a mixture (or a convex sum) of distributions corresponding to various $\Ga$, the convexity of entropy implies $\tS(t_1)\ge\tS(0)$.

At $t=t_1$, the mutual information between the state of the engine and that of the memory \footnote{
One easily finds that this is exactly equal to $\bar{I}$ in \rlb{Ibar}, which was defined as the mutual information between the engine at $t=0$ and the memory at $t=t_1$.
} is given by $\bar{I}=\{S(t_1)+\tS(t_1)\}-\bS(t_1)$.
By recalling that $\bS(t_1)=\bS(0)=S(0)+\tS(0)$ and $S(t_1)=S(0)$, we see that $\bar{I}=\tS(t_1)-\tS(0)$, i.e., the mutual information is equal to the increase of the entropy in the memory.

In the period $[t_1,t_2]$ of feedback, the entropy of the memory is preserved, and hence $\tS(t_2)=\tS(t_1)=\tS(0)+\bar{I}$.
The entropy of the engine can vary because there is a nontrivial feedback.
It may increase, decrease, or stay constant \footnote{
The argument which led to $\tS(t_1)\ge\tS(0)$ is no longer valid since the state of the memory at $t=t_1$ is correlated with the previous state of the engine.
}; the only constraint is the general inequality $\bS(t_2)\le S(t_2)+\tS(t_2)$.
By recalling that $\bS(t_2)=S(0)+\tS(0)$, this inequality is rewritten as
\eq
S(t_2)\ge S(0)-\bar{I},
\lb{SSI}
\en
which shows that the entropy of the engine may decrease but not more than by $\bar{I}$.
We can say that the mutual information $\bar{I}$ (generated during the measurement process) may be used as a resource to reduce the entropy of the engine (in the feedback process).
From \rlb{SSI} (which indeed is rigorous) and the nonnegativity of relative entropy one can rederive the generalized second law \rlb{2ndE} \footnote{%
Let $\bar{I}'':=\{S(t_2)+\tS(t_2)\}-\bS(t_2)$ be the mutual information between the engine and the memory in the final states.
Almost by definition we have $S(t_2)=S(0)-\bar{I}+\bar{I}''$, from which we get (again rigorously) an improved bound $\sbkt{\WE}\le(\bar{I}-\bar{I}'')/\beta$.
We do not know whether there is a corresponding integral fluctuation relation, but see \cite{ItoSagawa,HorowitzVaikuntanathan}.
}.
This is reasonable if we realize that the decrease in entropy by $\bar{I}$ is equivalent to the increase in the free energy by $\bar{I}/\beta$, which may be converted into work.

\midskip
\para{Derivation}%
Jarzynski relation \rlb{J} for the whole system is derived as in the original \cite{Jarzynski97} by noting that the time-evolution is one-to-one and measure-preserving.

We concentrate on the work \rlb{WE} of the engine.
Let $f(\Ga,\Up')$ be an arbitrary function of $\Ga$ and $\Up'$.
We find
\eqa
&\bkt{e^{\beta\WE(\Ga,\Up)}f(\Ga,\TmM(\Up))}
\nl&=
\int d\Ga d\Up\,e^{\beta\WE(\Ga,\Up)}f(\Ga,\TmM(\Up))\,\rhoz(\Ga,\Up)
\nl&=
\int d\Ga d\Up d\Up'\,\delta\bigl[\Up'-\TmM(\Up)\bigr]\,\rhoM(\Up)
\nl&\hspace{1cm}\times e^{\beta\WE(\Ga,\Up)}\,f(\Ga,\TmM(\Up))\,\rhoE(\Ga)
\nl
&=
\int d\Ga d\Up'\,\trho(\Up'|\Ga)\,e^{\beta\WE(\Ga,\Up)}
\,f(\Ga,\Up')\,\rhoE(\Ga),\notag
\intertext{%
where we used \rlb{Up'cond}.
Note that $\Up'$ is treated as a free variable here.
By substituting \rlb{can} and  \rlb{WE}, we get}
&=
\int d\Ga d\Up'\,\trho(\Up'|\Ga)\,f(\Ga,\Up')\,
\frac{e^{-\beta\HE(\TE(\Ga))}}{\ZE}.
\lb{eng1}
\ena
This is still a very complicated integral where the integrand depends nontrivially both on $\Ga$ and $\Up'$.
The integral becomes tractable if the integrand depends on $\Ga$ only through $\TE(\Ga)$.
This is possible in general only when one chooses
\eq
f(\Ga,\Up')=\frac{\nu(\Up')}{\trho(\Up'|\Ga)}
\lb{f}
\en
where $\nu(\Up')$ is arbitrary.
With this choice \rlb{eng1} becomes
\eqa
&\bkt{e^{\beta\WE}f}=
\int d\Ga d\Up'\,\nu(\Up')\,
\frac{e^{-\beta\HE(\TE(\Ga))}}{\ZE}
\nl&\hspace{0.4cm}
=\int d\Ga'' d\Up'\,\nu(\Up')\,
\frac{e^{-\beta\HE(\Ga'')}}{\ZE}=\int d\Up'\,\nu(\Up'),
\ena
where we have made the change of variable $\Ga''=\TE(\Ga)$, and used the Liouville theorem $d\Ga=d\Ga''$ (for each fixed $\Up'$).
We thus get $\bbkt{e^{\beta\WE}f}=1$ for $f$ given by \rlb{f} with an arbitrary $\nu(\Up')$ which satisfies $\int d\Up'\,\nu(\Up')=1$.

We next focus on the work \rlb{WM} done by the memory.
Let $g(\Ga,\Up')$ be an arbitrary function of $\Ga$ and $\Up'$.
Proceeding as in the derivation of the original Jarzynski relation \cite{Jarzynski97}, we have 
\eqa
&\bkt{e^{\beta\WM(\Ga,\Up)}g(\Ga,\TmM(\Up))}
\nl&=
\int d\Ga d\Up\,e^{\beta\WM(\Ga,\Up)}g(\Ga,\TmM(\Up))\,\rhoz(\Ga,\Up)
\nl&=
\int d\Ga d\Up\,g(\Ga,\TmM(\Up))\,\rhoE(\Ga)\,
\frac{e^{-\beta\HM(\TM(\Up))}}{\ZM}
\nl
&=
\int d\Ga d\Up''\,g(\Ga,(\TfM)^{-1}(\Up''))\,\rhoE(\Ga)\,
\frac{e^{-\beta\HM(\Up'')}}{\ZM},
\lb{mem1}
\ena
where we have made the change of variable $\Up''=\TM(\Up)$, and used the Liouville theorem $d\Up=d\Up''$ (for each fixed $\Ga$).
Again this is still a hardly tractable integral, but simplifies in general if $g$ is chosen to satisfy
\eq
\int d\Ga\,g(\Ga,\Up')\,\rhoE(\Ga)=1,
\lb{g}
\en
for any $\Up'$.
An obvious choice is $g=1$.
For $g$ satisfying \rlb{g}, the integral in \rlb{mem1} is easily evaluated and one gets $\bbkt{e^{\beta\WM}g}=1$.

To require $X=Y$ in \rlb{XY} corresponds to requiring $g=1/f$.
By substituting \rlb{f} into \rlb{g}, we find
\eq
1=\int d\Ga\,\frac{\trho(\Up'|\Ga)}{\nu(\Up')}\rhoE(\Ga)=\frac{\trho(\Up')}{\nu(\Up')},
\en
where we used \rlb{Up'prob}.
This uniquely determines $\nu(\Up')$ to be $\trho(\Up')$, and hence that
\eq
g(\Ga,\Up')=\frac{1}{f(\Ga,\Up')}=e^{I(\Ga,\Up')}.
\en

\midskip
\para{Discussion}%
As for a classical system operated by an outside agent, we have clarified which system should be called a Maxwell's demon in the most strict sense.
For such a system, we have established that the three relations \rlb{J}, \rlb{SUE}, and \rlb{SUM} form a unique triplet corresponding to the Sagawa-Ueda decomposition.
We believe that, as far as we concentrate on classical simple ``non-autonomous'' demons, these observations complete the project of Sagawa and Ueda to understand the essence of Maxwell's demon.

A remaining quite interesting challenge is to investigate whether similar results are possible for an ``autonomous Maxwell's demon'', a composite system which evolves under a fixed Hamiltonian without external operation \cite{MandalJarzynski,StrasbergSchallerBrandesEsposito1,MandalQuanJarzynski,StrasbergSchallerBrandesEsposito}.
It is likely that our criterion that ``the engine and the memory exchange only information'' may be realized only in certain limiting sense.
Even though such a criterion is expected to be quite useful in the analysis of demon-like engineering in nature (such as biological machines) or in the future technology.

\bigskip
It is a pleasure to thank Takahiro Sagawa, whom I regard almost as a coauthor, for discussions and suggestions which made the present work possible.
I also thank 
Takayuki Ariga, Sosuke Ito, and Shin-ichi Sasa
for useful discussions.



\newpage
\para{Appendix: Error-free system}%
Let us discuss the error-free version of the same problem of the engine and the memory.

We assume here that the state spaces are decomposed into disjoint unions as $\calE=\bigcup_{\mu=1}^m\calE_\mu$ and $\calM=\bigcup_{\mu=1}^m\calM_\mu$.
The time-evolution rule is basically the same.
But $\TmM$ now depends on $\Ga$ only through the unique index $\mu$ such that $\Ga\in\calE_\mu$, and hence is written as $\TmMmu$.
We assume that $\TmMmu$ is a one-to-one map from $\calM$ to $\calM_\mu$.
Thus the state $\Up'$ of the memory at $t=t_1$ specifies the index $\mu$ without any errors.
Likewise $\TfE$ now depends on $\Up'$ only through the unique $\mu'$ such that $\Up'\in\calM_{\mu'}$.
But since we already know that $\Up'\in\calM_\mu$, we have $\mu'=\mu$.
The time-evolution map is then denoted as $\TfEmu$, which is assumed to be a one-to-one map from $\TmE(\calE_\mu)$ to $\calE$.
The time-evolution maps for the whole interval is denoted as $\TEmu=\TfEmu\circ\TmE$ and $\TMmu=\TfM\circ\TmMmu$.

Again the map from $(\Ga,\Up)\in\calE\times\calM$ to $(\Ga'',\Up''):=(\calT_{\mu(\Ga)}(\Ga),\tilde{\calT}_{\mu(\Ga)}(\Up))\in\calE\times\calM$ is one-to-one and preservers the phase space volume.
We defined $\mu(\Ga)$ as the unique index such that $\Ga\in\calE_{\mu(\Ga)}$.

Let 
$
p_\mu:=\int_{\Ga\in\calE_\mu}\rhoE(\Ga)
$
be the probability that the state of the engine is initially in $\calE_\mu$.
Then we can show
\eqg
\bbkt{e^{\beta\{\WE(\Ga)+\WM(\Ga,\Up)\}}}=1,
\\
\bbkt{e^{\beta\WE(\Ga)+\log p_{\mu(\Ga)}}}=1,
\intertext{and}
\bbkt{e^{\beta\WM(\Ga,\Up)-\log p_{\mu(\Ga)}}}=1,
\eng
which are the Jarzynski relation and the two Sagawa-Ueda relations, respectively.
Note that we have the Shannon entropy function $-\log p_{\mu(\Ga)}$ instead of the mutual information function $I(\Ga,\Up')$.

\bigskip

Let us derive the Sagawa-Ueda relations, and also show the uniqueness of the Sagawa-Ueda decomposition.

First we concentrate on the time-evolution of the engine.
Then the only role of the memory is to ensure the correct feedback to the system.
For a fixed $\mu$, we have
\eq
\int_{\Ga\in\calE_\mu} d\Ga\,e^{\beta\WE(\Ga)}\rhoE(\Ga)
=
\int_{\Ga''\in\calE}d\Ga''\,\frac{e^{-\beta\HE(\Ga'')}}{\ZE}
=1,
\lb{eng2}
\en
where $\Ga''=\calT_\mu(\Ga)$ and we noted that $d\Ga=d\Ga''$.
Let $q_\mu$ be any quantity with $\sum_\mu q_\mu=1$.
Then by multiplying \rlb{eng2} by $q_\mu$ and summing up over $\mu$, one gets
\eq
\int_{\Ga\in\calE}d\Ga\,q_{\mu(\Ga)}\,e^{\beta\WE(\Ga)}\rhoE(\Ga)=1,
\en
which is nothing but $\bkt{e^{\beta\WE+\log q_\mu}}=1$.

Let us fix $\mu$, and examine the time-evolution of the memory.
It is convenient to define $\WM_\mu(\Up)=\HM(\Up)-\HM(\tilde{\calT}_\mu(\Up))$, which satisfies $\WM(\Ga,\Up)=\WM_{\mu(\Ga)}(\Up)$.
Then we get
\eq
\int d\Up\, e^{\beta\WM_\mu(\Up)}\rhoM(\Up)
=
\int_{\Up''\in\TfM(\calM_\mu)}d\Up''\,\frac{e^{-\beta\HM(\Up'')}}{\ZM}.
\en
Summing this over $\mu$ we get
\eq
\sum_\mu\int d\Up\, e^{\beta\WM_\mu(\Up)}\rhoM(\Up)
=1,
\en
which is rewritten as
\eq
\sum_{\mu}p_\mu\int d\Up\, \frac{1}{p_\mu}\,e^{\beta\WM_\mu(\Up)}\rhoM(\Up)
=1.
\en
This is nothing but the desired Sagawa-Ueda relation $\bbkt{e^{\beta\WM-\log p_\mu}}=1$.
Interestingly the fluctuation relation is essentially unique in this situation.
From the requirement corresponding to $X=Y$, we uniquely determine $q_\mu$ to be $p_\mu$.

\end{document}